\documentclass[conference]{IEEEtran}
\IEEEoverridecommandlockouts
\usepackage{cite}
\usepackage{overpic}
\usepackage{amsmath,amssymb,amsfonts}
\usepackage{graphicx}
\usepackage{textcomp}
\usepackage{xcolor}
\usepackage{float}
\usepackage{amsthm}
\usepackage{graphicx}
\usepackage{epstopdf}
\usepackage{amsmath,bm}
\usepackage{amsfonts}
\usepackage{amssymb}
\usepackage{color}
\usepackage{multirow}
\usepackage{multicol}
\usepackage{soul,xcolor}
\usepackage{algorithm}
\usepackage{algpseudocode}%

\newcommand{\condSum}[3]{\overset{#3}{\underset{\underset{#2}{#1}}{\sum}}}

\newcommand{\mathacr}[1]{\mathsf{#1}}
\theoremstyle{plain}

\newtheorem{lemma}{Lemma}

\newcommand{\vect}[1]{\mathbf{#1}}

\def\diag{\mathrm{diag}}

\def\Htran{\mbox{\tiny $\mathrm{H}$}}
\def\Ttran{\mbox{\tiny $\mathrm{T}$}}
\def\CN{\mathcal{N}_{\mathbb{C}}} 
\def\imagunit{\mathsf{j}} 
\def\BibTeX{{\rm B\kern-.05em{\sc i\kern-.025em b}\kern-.08em
    T\kern-.1667em\lower.7ex\hbox{E}\kern-.125emX}}
\begin{document}

\title{RIS-Assisted Massive MIMO with Multi-Specular Spatially Correlated Fading \vspace{-0.5cm}
\thanks{This work was supported by the FFL18-0277 grant from the Swedish Foundation for Strategic Research.}
}

\author{\IEEEauthorblockN{\"Ozlem Tu\u{g}fe Demir\IEEEauthorrefmark{1} and Emil Bj\"ornson\IEEEauthorrefmark{1}\IEEEauthorrefmark{2}}
	\IEEEauthorblockA{{\IEEEauthorrefmark{1}Department of Computer Science, KTH Royal Institute of Technology, Kista, Sweden, \{ozlemtd, emilbjo\}@kth.se
		} \\ {\IEEEauthorrefmark{2}Department of Electrical Engineering, Link\"oping University, Link\"oping, Sweden
		} 
	}
	
}

\maketitle

\begin{abstract}
Reconfigurable intelligent surfaces (RISs) have attracted great attention as a potential beyond 5G technology. These surfaces consist of many passive elements of metamaterials whose impedance can be controllable to change the phase, amplitude, or other characteristics of wireless signals impinging on them. Channel estimation is a critical task when it comes to the control of a large RIS when having a channel with a large number of multipath components. In this paper, we propose a novel channel estimation scheme that exploits spatial correlation characteristics at both the massive multiple-input multiple-output (MIMO) base station and the planar RISs, and other statistical characteristics of multi-specular fading in a mobile environment. Moreover, a novel heuristic for phase-shift selection at the RISs is developed, inspired by signal processing methods that are effective in conventional massive MIMO. Simulation results demonstrate that the proposed uplink RIS-aided framework improves the spectral efficiency of the cell-edge mobile users substantially in comparison to a conventional single-cell massive MIMO system.%
\end{abstract}
\vspace{-2mm}
\section{Introduction}
\vspace{-1.5mm}
Reconfigurable intelligent surfaces (RISs), also known as intelligent reflecting surfaces \cite{Wu2019}, are envisioned as a key technology in shaping the wireless medium by software-controlled reflection of the individual  propagation paths with an aim to boost the desired signal power at the receiver. The RISs are envisaged to be deployed as planar surfaces on the facades, walls, or ceilings of the buildings and consist of a large number of reflecting elements \cite{ RIS_Renzo_JSAC}. Each element can be made of metamaterial, acts as an isotropic scatterer when it is sub-wavelength-sized, and the impedance can be tuned to create a phase-shift pattern over the surface that reflects an incident wave as a beam in a desirable direction \cite{emil_rayleigh_fading_ris}.

Massive multiple-input multiple-output (MIMO) is the 5G technology that allows serving several user equipments (UEs) on the same time-frequency resources by spatial multiplexing \cite{massivemimobook}. The performance of massive MIMO when it is combined with the new RIS technology has been studied in  \cite{RISmassiveMIMO_perfect_CSI,RISmassiveMIMO_overhead,RISmassiveMIMO_LOS}. In the existing works, either perfect channel state information (CSI) is assumed, only line-of-sight (LOS) components, or previous realizations of the channels are used to optimize RIS phase-shift design. In this paper, we go beyond that by considering two important tasks for RIS-assisted uplink massive MIMO: i) A channel estimation with simple and low training overhead; ii) phase-shift and receive combiner design based on channel estimates without resorting to complex optimization. 

Estimation of all the individual channel coefficients in an RIS-aided wireless network requires a huge number of pilot symbols since each channel from a UE to the base station (BS) through a single element of an RIS should effectively be estimated to select the phase-shifts properly in the data transmission phase. Channel estimation for RISs is a key open problem \cite{RIS_emil_magazine} and there are some initial studies for this topic in different contexts \cite{RISchannelEstimation_nested_cascaded,	RISchannelEstimation_nested_LS_single,RISchannelEstimation_nested_cascaded2,
		RISchannelEstimation_nested_iterative,
	RISchannelEstimation_nested_knownBSRIS,
	RISchannelEstimation_nested_knownBSRIS2}. Note that the existing works either do not exploit the spatial correlation between BS antennas and RIS elements \cite{RISchannelEstimation_nested_cascaded,RISchannelEstimation_nested_LS_single}, and/or they are not intended for massive MIMO  \cite{RISchannelEstimation_nested_cascaded2,RISchannelEstimation_nested_LS_single,RISchannelEstimation_nested_iterative}. Some of them assume that the BS-RIS channel is deterministic \cite{RISchannelEstimation_nested_knownBSRIS2,RISchannelEstimation_nested_knownBSRIS}. Different from these works, our main contributions are as follows:
	 \vspace{-1.5mm}
\begin{itemize}
	\item For the first time, we consider channel estimation in an RIS-aided system by taking both the spatial correlation at the BS and multiple RISs, and random phase-shifts on the specular (dominant) paths, which are usually neglected in the literature, into account.
	\item We derive a Bayesian channel estimator for all the individual propagation paths by using a structured pilot assignment with a low training overhead.
	\item We propose a low-complexity closed-form phase-shift selection scheme at the RISs and provide a respective achievable uplink spectral efficiency.
	\end{itemize}

\vspace{-3mm}
\section{System Model}
\vspace{-2.2mm}
We consider an uplink single-cell massive MIMO system that is assisted by multiple RISs. The BS has $M$ antennas and each RIS has $N$ reflecting elements. The number of RISs and UEs is $L$ and $K$, respectively. Each UE is equipped with a single antenna. To achieve low complexity in the design of the phase-shifts of the RIS elements, we assume that each one is assigned to one UE and its phase-shift is selected to maximize the effective channel strength of that UE. Let $N_k\geq 0$ denote the number of reflecting elements assigned to UE $k$, for $k=1,\ldots,K$ and we have $\sum_{k=1}^KN_k=LN$ if all the RIS reflecting elements are assigned to the UEs. Throughout the communication, the BS first estimates the channels and optimizes the phase-shifts. Then, it sends the phase-shift information to the RISs in each coherence block.

We consider the conventional block fading model where the channel responses in each coherence block are frequency-flat and time-invariant, thus  represented by fixed complex scalars. We let $\tau_c$ denote the total number of samples per coherence block. Each coherence block is divided into two phases: uplink training  and uplink data transmission with $\tau_p$ and $\tau_c-\tau_p$ channel uses, respectively.

Let $\vect{h}_{k}\in \mathbb{C}^{M}$  denote the channel from UE $k$ to the BS. Generalizing the spatially correlated Rician fading model \cite{DemirICASSP20} to consider an arbitrary number of specular components, each channel realization is expressed as
\vspace{-2.3mm}
\begin{equation} \label{eq:channel-UE-BS}
 \vect{h}_{k}=\sum_{s=1}^{S_k^{\mathrm{h}}}e^{\imagunit\theta^{\mathrm{h}}_{k,s}}\bar{\vect{h}}_{k,s}+\tilde{\vect{h}}_k
 \vspace{-2mm}
\end{equation}
where the vector $\bar{\vect{h}}_{k,s}$ is the array steering vector scaled by the square root of the corresponding link gain for the $s$th specular component of the channel. $S_k^{\mathrm{h}}$ is the number of specular components, which are also called ``dominant'' paths. The nonspecular part of the channel $\tilde{\vect{h}}_k$ represents the summation of other diffusely propagating multipath components. The vectors $\bar{\vect{h}}_{k,s}$ are fixed for a given setup. If there is an LOS path between the BS and UE $k$, then one of the specular components $e^{\imagunit\theta^{\mathrm{h}}_{k,s}}\bar{\vect{h}}_{k,s}$ corresponds to this path. Note that microscopic movements induce random phase-shifts on each individual propagation path and the one that affects the $s$th specular component of the channel is denoted by  $\theta^{\mathrm{h}}_{k,s}$. These phase-shifts change from coherence block to coherence block. Hence, they are not known in advance and are modeled as an independent uniform random variable in $[0,2\pi)$, i.e., $\theta^{\mathrm{h}}_{k,s} \sim \mathcal{U}[0,2\pi)$. The nonspecular paths are subject to similar phase-shifts, which give rise to small-scale fading modeled by a Gaussian distribution $\tilde{\vect{h}}_k\sim \CN(\vect{0}_M,\vect{R}^{\mathrm{h}}_k)$ that also takes an independent realization in each coherence block. The matrix  $\vect{R}^{\mathrm{h}}_k\in \mathbb{C}^{M \times M}$ describes the spatial correlation between the channel realizations observed at different BS antennas and other long-term channel effects such as geometric pathloss and shadowing. These matrices are also fixed for a given setup and can, thus, be assumed to be known.

 The channel from UE $k$ to RIS $\ell$ is denoted by 
 $\vect{f}_{k\ell}\in \mathbb{C}^{N}$ and using the same multi-specular correlated fading model as in \eqref{eq:channel-UE-BS}, it is expressed as
 \vspace{-2.4mm}
 \begin{equation} \label{eq:channel-UE-IRS}
  \vect{f}_{k\ell}=\sum_{s=1}^{S_{k\ell}^{\mathrm{f}}}e^{\imagunit\theta^{\mathrm{f}}_{k\ell,s}}\bar{\vect{f}}_{k\ell,s}+\tilde{\vect{f}}_{k\ell}
  \vspace{-1.7mm}
 \end{equation}
 where $\theta^{\mathrm{f}}_{k\ell,s} \sim \mathcal{U}[0,2\pi)$ and $\tilde{\vect{f}}_{k\ell}\sim \CN(\vect{0}_{N},\vect{R}^{\mathrm{f}}_{k\ell})$.  We will denote the channel from UE $k$ to the $N_i$ reflecting elements that are assigned to UE $i$ by ${\vect{f}}^{\prime}_{ki}\in \mathbb{C}^{N_i}$ and it is constructed by the elements of the vectors $\vect{f}_{k\ell}$, for $\ell=1,\ldots,L$ at the corresponding indices.

 Note that the BS and the RISs are usually stationary. Furthermore, in a well-designed system, RISs would be deployed to have LOS paths to the BS (see \cite[Fig.~7]{Bjornson2021}). Hence, we will not consider a random phase-shift on the LOS component and express the channel between the BS and RISs by separating the LOS part. However, there can be changes in the non-line-of-sight (NLOS) paths due to the time-varying environment. Hence, we can model the channel $\vect{G}_{\ell} \in \mathbb{C}^{M\times N}$ as
  \vspace{-2mm}
\begin{equation}\label{eq:channel-BS-IRS}
\vect{G}_{\ell} = \bar{\vect{G}}_{\ell,1} + \sum_{s=2}^{S_{\ell}^{\mathrm{G}}}e^{\imagunit\theta^{\mathrm{G}}_{\ell,s}}\bar{\vect{G}}_{\ell,s}+\tilde{\vect{G}}_{\ell}
\vspace{-1.7mm}
\end{equation}
which is the channel from RIS $\ell$ to the BS. Here $\bar{\vect{G}}_{\ell,1}$ is the LOS part and the second term includes the $S_{\ell}^{\mathrm{G}}-1$ specular components with random phase-shifts $\theta^{\mathrm{G}}_{\ell,s} \sim\mathcal{U}[0,2\pi)$. Using the Kronecker model \cite{Shiu2000a} with the receive and transmit correlation matrices for the BS and RISs, $\vect{R}_{\ell}^{\mathrm{G,BS}}\in \mathbb{C}^{M\times M}$ and $\vect{R}_{\ell}^{\mathrm{G,RIS}}\in \mathbb{C}^{N \times N}$, for $\ell=1,\ldots,L$, respectively, the nonspecular part of the channel $\vect{G}_{\ell}$ can be expressed as
\vspace{-2mm}
\begin{equation}
\tilde{\vect{G}}_{\ell} = \left(\vect{R}_{\ell}^{\mathrm{G,BS}}\right)^{\frac12}\vect{W}_{\ell}\left(\vect{R}_{\ell}^{\mathrm{G,RIS}}\right)^{\frac12} \label{eq:Gtilde}
\vspace{-1.7mm}
\end{equation}
where the elements of $\vect{W}_{\ell}\in \mathbb{C}^{M \times N}$ are independent and identically distributed (i.i.d.) standard complex Gaussian random variables. Let $\vect{G}^{\prime}_k\in \mathbb{C}^{M\times N_k}$ denote the channel from the subset of RIS elements assigned to UE $k$ to the BS. This matrix is constructed by picking the columns of the matrices $\vect{G}_{\ell}$, for $\ell=1,\ldots,L$ corresponding to the assigned RIS elements.

During the uplink transmission phase, the received signal at the BS can be expressed as
\vspace{-2.2mm}
\begin{equation}\label{eq:received_signal} \vect{y}=\sum_{i=1}^K\left(\vect{h}_i+\sum_{j=1}^K\vect{G}_j^{\prime}\vect{\Phi}_j\vect{f}^{\prime}_{ij}\right)s_i+\vect{n} 
\vspace{-1.7mm}
\end{equation}
where
$s_i$ is either the pilot or data signal of UE $i$ and $\vect{n}\sim\CN(\vect{0}_M,\sigma^2\vect{I}_M)$ is the additive noise. The phase-shift of the $n$th element of the $j$th RIS subset (assigned to UE $j$) is represented by $\phi_{j,n}\in \mathbb{C}$ with the unit-modulus constraint $\left\vert \phi_{j,n} \right\vert=1$, which is the $(n,n)$th element of the diagonal matrix $\vect{\Phi}_{j}\in \mathbb{C}^{N_j \times N_j}$. We consider a setup with continuous phase control, which is fully possible in practice \cite{RISfieldtrial}.
Defining
\vspace{-2.2mm}
\begin{equation} \label{eq:cascaded}
\vect{H}^{\prime}_{ij}\triangleq \vect{G}^{\prime}_{j}\diag(\vect{f}^{\prime}_{ij}) \in \mathbb{C}^{M \times N_j},
\vspace{-1.7mm}
\end{equation}  
where $\diag(\vect{f}^{\prime}_{ij})$ denotes the $N_j\times N_j$ diagonal matrix with entries as the elements of the vector $\vect{f}^{\prime}_{ij}$, the received signal in \eqref{eq:received_signal} can also be expressed in the following form
\vspace{-2mm}
\begin{equation}\label{eq:received_signal1b} \vect{y}=\sum_{i=1}^K\left(\vect{h}_i+\sum_{j=1}^K\vect{H}^{\prime}_{ij}\bm{\phi}_j\right)s_i+\vect{n} 
\vspace{-1.5mm}
\end{equation}
where $\bm{\phi}_j\in \mathbb{C}^{N_j}$ is the vector whose $n$th element is $\phi_{j,n}$.

\vspace{-1.3mm}
\section{Phase-Shift Selection}\label{sec:phase}
\vspace{-1mm}

In this section, we will introduce a phase-shift selection scheme that does not require any complex optimization tool or algorithm and, hence, can be implemented per coherence block by using the estimates for each channel realization. In the uplink payload data transmission phase, the received signal at the BS is given as in \eqref{eq:received_signal} or \eqref{eq:received_signal1b} with
$s_i$ being the uplink data signal of UE $i$ with transmit power $p_i$, i.e., $\mathbb{E}\{\left\vert s_i\right\vert^2\}=p_i$. Let $\widehat{\vect{h}}_i$ and $\widehat{\vect{H}}^{\prime}_{ij}$ denote the estimates of the direct BS-UE channel $\vect{h}_i$ and the cascaded BS-RIS-UE channel $\vect{H}^{\prime}_{ij}$ in \eqref{eq:received_signal1b}. Since the $j$th RIS subset is assigned to serve UE $j$, in the proposed scheme, $\phi_{j,n}$ will be selected to maximize the channel strength of UE $j$. More precisely, considering only the cascaded channel through the $j$th RIS subset in \eqref{eq:received_signal1b}, our aim is to maximize the norm of the respective portion of the overall estimated channel, i.e., $\widehat{\vect{h}}_j+\widehat{\vect{H}}^{\prime}_{jj}\bm{\phi}_j$. Under the unit modulus constraints on the elements of $\bm{\phi}_j$, the corresponding optimization problem is given as
\vspace{-2mm}
\begin{equation}
\begin{aligned}[b] 
&\underset{\bm{\phi}_j}{\textrm{maximize}} \quad \left\Vert \widehat{\vect{h}}_j+\widehat{\vect{H}}^{\prime}_{jj}\bm{\phi}_j\right\Vert^2  \\
&\textrm{subject to} \quad \vert\phi_{j,n}\vert=1, \quad n=1,\ldots,N_j. \label{eq:norm-maximization}\end{aligned}
\end{equation}
The problem \eqref{eq:norm-maximization} is non-convex but semidefinite programming (SDP) with rank relaxation can be used to obtain a suboptimal solution similar to the single-UE problem in \cite[Sec.~III-A]{Wu2019}. SDP is computationally infeasible when having a massive number of BS antennas and RIS elements. A heuristic approximation to the optimal solution of \eqref{eq:norm-maximization} can be obtained by relaxing the unit modulus constraints:
\vspace{-2.2mm}
\begin{equation}
\begin{aligned}[b] 
&\underset{\bm{\phi}_j}{\textrm{maximize}} \quad \left\Vert \widehat{\vect{h}}_j+\widehat{\vect{H}}^{\prime}_{jj}\bm{\phi}_j\right\Vert^2  \\
&\textrm{subject to} \quad \Vert \bm{\phi}_j\Vert^2\leq N_j. \label{eq:norm-maximization3}\end{aligned}\vspace{-2mm}
\end{equation}
The above problem is still non-convex but it can be solved optimally using eigenvalue decomposition and a bisection search for root finding as shown in the following lemma.
\vspace{-1.7mm}
\begin{lemma}\label{lemma:norm_maximization}
	Let $\lambda_{j,d}\geq0$ be the nonnegative eigenvalues and $\vect{u}_{j,d}\in \mathbb{C}^{N_j}$ be the corresponding orthonormal eigenvectors of $(\widehat{\vect{H}}^{\prime}_{jj})^{\Htran}\widehat{\vect{H}}^{\prime}_{jj}$.
If $\widehat{\vect{h}}_j=\vect{0}_{N_j}$, the optimal solution to the problem \eqref{eq:norm-maximization3} is given by $\bm{\phi}_j^{\star}=\sqrt{N_j}\vect{u}_{j,\bar{d}}$ where $\bar{d}$ is the index corresponding to the dominant eigenvector. Otherwise, the optimal solution is  
\vspace{-2mm}
\begin{equation}
\bm{\phi}_j^{\star} = \sum_{d=1}^{N_j} \frac{\vect{u}_{j,d}\vect{u}_{j,d}^{\Htran}(\widehat{\vect{H}}^{\prime}_{jj})^{\Htran}\widehat{\vect{h}}_j}{\gamma^{\star}-\lambda_{j,d}}, \quad d=1,\ldots,N_j, \label{eq:phi}
\vspace{-2mm}
\end{equation}
where $\gamma^{\star}>\max_d\lambda_{j,d}$ is the unique root of
\vspace{-2mm}
\begin{equation}
\sum_{d=1}^{N_j}\frac{\left\vert\vect{u}_{j,d}^{\Htran}(\widehat{\vect{H}}^{\prime}_{jj})^{\Htran}\widehat{\vect{h}}_j\right\vert^2}{\left(\gamma-\lambda_{j,d}\right)^2}=N_j.\label{eq:root2}
\vspace{-2mm}
\end{equation} 
\begin{proof} The proof is omitted due to space limitation.
	\end{proof}
\end{lemma}
\vspace{-2mm}

We select the RIS phase-shifts by picking the phase-shifts of the optimal $\bm{\phi}_j^{\star}$ given in Lemma~\ref{lemma:norm_maximization}, i.e.,
\vspace{-2mm}
\begin{equation}\label{eq:phase-shift1}
\phi_{j,n} = e^{\imagunit\angle \phi_{j,n}^{\star} }, \quad n=1,\ldots,N_j, \quad j=1,\ldots,K.
\vspace{-2mm}
\end{equation}

 Note that selecting the phase-shifts based on more complex optimization algorithms as in the previous works \cite{Wu2019,RISchannelEstimation_nested_knownBSRIS2} will necessitate solving a non-convex problem with computationally time-consuming algorithms. Moreover, a new optimization problem should be solved in each coherence block, which is impractical in mobile scenarios with ms-range coherence times. The proposed scheme in \eqref{eq:phase-shift1} does not require any heavy processing and, hence, it is computationally efficient with the same order as the linear receive combining schemes of conventional massive MIMO systems. 
 
 \vspace{-1mm}

\section{Uplink Spectral Efficiency}\label{sec:SE}
\vspace{-1mm}

Let $\vect{v}_k\in \mathbb{C}^M$ denote the receive combining vector that is applied to the received signal $\vect{y}$ given in \eqref{eq:received_signal} and \eqref{eq:received_signal1b} for the decoding of the UE $k$'s uplink data at the BS. Defining
\vspace{-2mm}
\begin{equation} \label{eq:bk}
\vect{b}_k=\vect{h}_k+\sum_{j=1}^K\vect{G}_j^{\prime}\vect{\Phi}_j\vect{f}^{\prime}_{kj}=\vect{h}_k+\sum_{j=1}^K\vect{H}^{\prime}_{kj}\bm{\phi}_j, \vspace{-1.5mm}
\end{equation}
for $k=1,\ldots,K$, the BS obtains the soft estimate of $s_k$ as
\vspace{-2mm}
\begin{equation}
\widehat{s}_k=\vect{v}_k^{\Htran}\vect{y}=\vect{v}_k^{\Htran}\vect{b}_ks_k
+\condSum{i=1}{i\neq k}{K}\vect{v}_k^{\Htran}\vect{b}_is_i+\vect{v}_k^{\Htran}\vect{n}\label{eq:soft_estimate}.
\vspace{-2mm}
\end{equation}
An ergodic spectral efficiency (SE) of the proposed RIS-assisted massive MIMO system can be obtained as follows.

\vspace{-1.3mm}
\begin{lemma} \label{eq:capacity1}
	An achievable SE of UE $k$ is
	\vspace{-2mm}
	\begin{equation} \label{eq:rate-expression-general}
	\mathrm{SE}_{k} = \frac{\tau_c-\tau_p}{\tau_c} \log_2  \left( 1 + \mathrm{SINR}_{k}   \right) \quad \textrm{bit/s/Hz}
	\vspace{-2mm}
	\end{equation}
	where the effective signal-to-interference-plus-noise ratio (SINR) is given by
	\vspace{-2mm}
	\begin{align} \label{eq:SINR}
	&\mathrm{SINR}_{k} = \nonumber\\
&	\frac{ p_{k} \left |\mathbb{E} \left\{  \vect{v}_{k}^{\Htran} \vect{b}_k \right\}\right|^2  }{ 
		\sum\limits_{i=1}^K p_{i}  \mathbb{E} \left\{| \vect{v}_{k}^{\Htran} \vect{b}_{i} |^2\right\}
		- p_{k} \left |\mathbb{E} \left\{  \vect{v}_{k}^{\Htran} \vect{b}_{k} \right\}\right|^2  + \sigma^2 \mathbb{E} \left\{\|  \vect{v}_{k} \|^2\right\}
	}.
	\end{align}
\end{lemma}

\vspace{-2mm}

\begin{proof}
	The SE in \eqref{eq:rate-expression-general} is obtained by treating  \eqref{eq:soft_estimate} as an interference channel with the known channel response $\mathbb{E}\left\{\vect{v}_k^{\Htran}\vect{b}_k\right\}$ and the interference $\vect{v}_k^{\Htran}\vect{y}-\mathbb{E}\left\{\vect{v}_k^{\Htran}\vect{b}_k\right\}s_k$. Then, using \cite[Corollary~1.3]{massivemimobook}, we obtain the given result.
\end{proof}
\vspace{-1.5mm}

Note that Lemma~\ref{eq:capacity1} is valid for any selection of receive combining vector $\vect{v}_k$. Maximum ratio (MR) combining can be used to maximize the received power, which for the considered RIS-assisted massive MIMO system can be defined as
\vspace{-1.5mm}
\begin{equation}\label{eq:MR}
\vect{v}_k^{\mathrm{MR}}= \widehat{\vect{b}}_k
\vspace{-2mm}
\end{equation}
where $\widehat{\vect{b}}_k$ is the estimate of the overall channel $\vect{b}_k$ obtained by the proposed estimator that will be described in the following section. To suppress interference, we can use regularized zero-forcing (RZF)  with 
\vspace{-2mm}
\begin{equation}\label{eq:RZF}
\vect{v}_k^{\mathrm{RZF}}=\left(\sum_{i=1}^Kp_i\widehat{\vect{b}}_i\widehat{\vect{b}}_i^{\Htran}+\sigma^2\vect{I}_M\right)^{-1}\widehat{\vect{b}}_k.
\end{equation}

\vspace{-1mm}

\section{Channel Estimation} \label{sec:individual}
\vspace{-1mm}

To select the RIS phase-shifts in \eqref{eq:phase-shift1}, we need to estimate the direct channels $\vect{h}_i$ and the cascaded RIS channels $\vect{H}^{\prime}_{ij}$ in each coherence block and use those estimates in the phase-shift and receive combiner selection.

For the proposed channel estimation scheme, the number of samples that are allocated for pilot transmission in the training phase is  $\tau_p=(LR+1)K$ under the assumption that $\tau_p<\tau_c$. Here, $1\leq R\leq N$ is a predefined integer parameter such that $N/R$ is also an integer. $R$ is equal to the number of non-overlapping RIS sub-surfaces with $N/R$ elements in a given RIS, which are introducing the same phase-shifts in the pilot training phase. In this way, the required length of the orthogonal pilot sequences is reduced from $(LN+1)K$, which is a huge number when having a large number $LN$ of RIS elements, to $(LR+1)K$. For the proposed method, the parameter $R$ can be adjusted arbitrarily enabling lower training overhead. The value of $R$ for a given setup can be adjusted by considering the trade-off between the pilot length and the channel estimation performance.

\begin{figure}[t]
\vspace{2mm}
	\hspace{1cm}
\centering
		\begin{overpic}[width=6.8cm,tics=10]{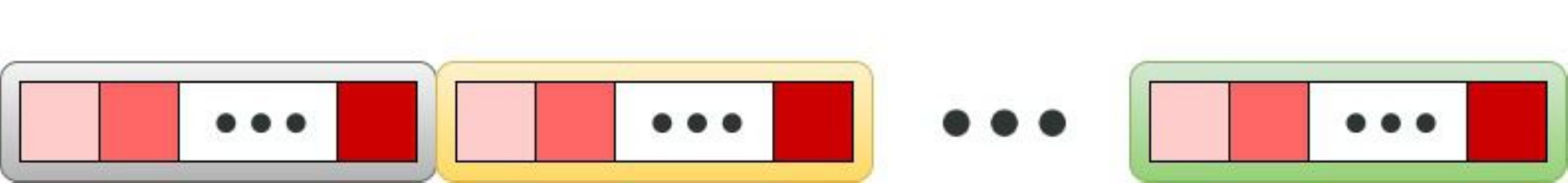}
			\put(-19.5,9.5){\small RIS $(\ell,r)$}
			\put(-19.5,-14){\small Received}
			\put(-22,-18.5){\small pilot signals}
			\put(7,10){ $\psi_{\ell r,0}$}
			\put(35,10){ $\psi_{\ell r,1}$}
			\put(77.5,10){ $\psi_{\ell r,LR}$}
			\put(-16,-5){\small UE $k$}
			\put(9.5,-4.5){ $\bm{\varphi}_k$}
			\put(11.5,-16){\small $\vect{Y}^{\mathrm{p}}_{0}$}
			\put(39,-16){\small $\vect{Y}^{\mathrm{p}}_{1}$}
			\put(83,-16){\small $\vect{Y}^{\mathrm{p}}_{LR}$}			
			\put(37,-4.5){ $\bm{\varphi}_k$}
			\put(81.5,-4.5){ $\bm{\varphi}_k$}
			\put(1.2,-9){\small$\varphi_{k,1} \ldots\varphi_{k,K}$}
			\put(29.4,-9){\small$\varphi_{k,1} \ldots\varphi_{k,K}$}
			\put(73.5,-9){\small$\varphi_{k,1} \ldots\varphi_{k,K}$}
		\end{overpic}
	\vspace{9mm}
	\caption{Pilot structure and RIS phase-shifts. }
	\label{fig:channel_estimation}
	\vspace{-1mm}
\end{figure}
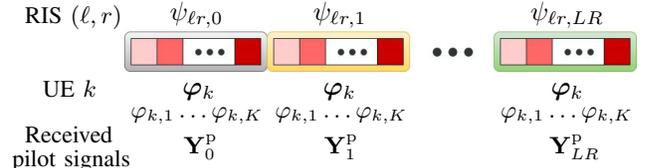

We consider $LR+1$ pilot transmission intervals each of which spans $K$ channel uses as depicted in Fig.~\ref{fig:channel_estimation}. Let $\bm{\varphi}_k\in \mathbb{C}^{K}$ denote the pilot signal assigned to UE $k$ for each training interval where $\Vert\bm{\varphi}_{k}\Vert^2=1$, $\forall k$. The pilot signals are mutually orthogonal between UEs, i.e., $\bm{\varphi}_k^{\Htran}\bm{\varphi}_i=0$, $\forall i\neq k$. We will use the index $t=0$ for the first training interval and $t=1,\ldots,LR$ for the consecutive $LR$ intervals. The phase-shift introduced by each element in the $r$th sub-surface of RIS $\ell$ in the $t$th pilot interval is represented by $\psi_{\ell r,t}\in\mathbb{C}$ with $\vert \psi_{\ell r,t}\vert=1$, for $\ell=1,\ldots,L$, $r=1,\ldots,R$, and $t=0,\ldots,LR$. Let RIS $(\ell,r)$ refer to the elements in the $r$th sub-surface of RIS $\ell$ as in Fig.~\ref{fig:channel_estimation}. Let also $\bm{\psi}_{\ell r}\in \mathbb{C}^{LR+1}$ denote the vector that is constructed by the phase-shifts of the RIS $(\ell,r)$ throughout the whole training interval, i.e., $\bm{\psi}_{\ell r}=[\psi_{\ell r,0} \ \ldots \ \psi_{\ell r,LR}]^{\Ttran}\in \mathbb{C}^{LR+1}$, for $\ell=1,\ldots,L$ and $r=1,\ldots,R$. We select them so that they are mutually orthogonal and also orthogonal to the all ones vector ${\bf 1}_{LR+1}=[ 1\ \ldots\ 1]^{\Ttran}\in \mathbb{C}^{LR+1}$, i.e.,  $\bm{\psi}_{\ell r}^{\Htran}\bm{\psi}_{\ell^{\prime}r^{\prime}}=0$, $\forall (\ell^{\prime},r^{\prime}) \neq (\ell,r)$ and $\bm{\psi}_{\ell r}^{\Htran}{\bf 1}_{LR+1}=0$, for  $\forall \ell, \forall r$.  This is satisfied by the $LR$ columns of the $(LR+1)\times (LR+1)$ discrete Fourier transform (DFT) matrix by excluding the first column. To make the notation simpler, let us first define 
 \vspace{-2mm}
\begin{equation} \label{eq:cascaded2}
\vect{H}_{k\ell}\triangleq \vect{G}_{\ell}\diag(\vect{f}_{k\ell}) \in \mathbb{C}^{M \times N}
\vspace{-2mm}
\end{equation}  
in analogy with the definition in \eqref{eq:cascaded}. The columns of the matrix $\vect{H}_{k\ell}$ correspond the cascaded channels from UE $k$ to the BS through RIS $\ell$. Let also $\mathrm{RIS}_{\ell,r}$ denote the set of element indices corresponding to the RIS $(\ell,r)$. The channels $[\vect{H}_{k\ell}]_{:n}$ for $n\in \mathrm{RIS}_{\ell,r}$ will be affected by the same phase-shift $\psi_{\ell r,t}$ in the $t$th training interval. Using these definitions, the received signal at the BS in the $t$th interval is thus given as
\vspace{-2mm}
\begin{equation}\label{eq:received_pilot}
\vect{Y}^{\mathrm{p}}_{t}\!\!=\!\!\sum_{i=1}^K\!\!\sqrt{K\eta}\!\!\left(\!\vect{h}_i\!+\!\sum_{\ell=1}^{L}\sum_{r=1}^{R} \psi_{\ell r,t}\!\!\!\!\sum_{n\in \mathrm{RIS}_{\ell,r}}\!\!\!\!\!\left[\vect{H}_{i\ell}\right]_{:n}\!\right)\bm{\varphi}_{i}^{\Ttran}\!+\!\vect{N}^{\mathrm{p}}_{t}\vspace{-1.5mm}
\end{equation}
where $\vect{N}_t^{\mathrm{p}}\in \mathbb{C}^{M\times K}$ is the additive noise with i.i.d. $\CN(0,\sigma^2)$ elements. The pilot transmit power is denoted by $\eta$.

After correlating   $\vect{Y}^{\mathrm{p}}_t$ with $\bm{\varphi}_{k}$, we obtain the sufficient statistics for the estimation of $\vect{h}_k$ and $\vect{H}_{k\ell}$, for $\ell=1,\ldots,L$; 
\vspace{-2.2mm}
\begin{align}
\vect{z}_{k,t}^{\mathrm{p}}&\!=\!\vect{Y}^{\mathrm{p}}_t\bm{\varphi}_{k}^*\!\nonumber\\
&\!=\!\sqrt{K\eta}\!\left(\!\vect{h}_k\!+\!\sum_{\ell=1}^{L}\sum_{r=1}^{R}\psi_{\ell r,t}\!\!\!\sum_{n\in \mathrm{RIS}_{\ell,r}}\!\left[\vect{H}_{k\ell}\right]_{:n}\!\right)\!+\!\tilde{\vect{n}}_{k,t}^{\mathrm{p}} \label{eq:sufficient_statistics_k}
\end{align}
where $\tilde{\vect{n}}_{k,t}^{\mathrm{p}}=\vect{N}_t^{\mathrm{p}}\bm{\varphi}_{k}^*\sim\CN(\vect{0}_M,\sigma^2\vect{I}_{M})$. Since ${\bf 1}_{LR+1}$ is orthogonal to the vectors $\bm{\psi}_{\ell r}$, we have ${\bf 1}_{LR+1}^{\Htran}\bm{\psi}_{\ell r}= \sum_{t=0}^{LR}\psi_{\ell r,t}=0$, for $\ell=1,\ldots,L$ and $r=1,\ldots,R$. We can obtain the sufficient statistics for the estimation of direct channel $\vect{h}_k$ using all the received pilot signals:
\vspace{-1.5mm}
\begin{equation}
\vect{z}_{k}^{\mathrm{p,h}}=\frac{\sum_{t=0}^{LR}\vect{z}_{k,t}^{\mathrm{p}}}{\sqrt{LR+1}}=\sqrt{K(LR+1)\eta}\vect{h}_k+\tilde{\vect{n}}_{k}^{\mathrm{p,h}} \label{eq:sufficient_statistics_k_0}
\vspace{-1.5mm}
\end{equation}
where $\tilde{\vect{n}}_{k}^{\mathrm{p,h}}=\sum_{t=0}^{LR}\tilde{\vect{n}}_{k,t}^{\mathrm{p}}\big/\sqrt{LR+1} \sim\CN(\vect{0}_M,\sigma^2\vect{I}_{M})$. Note that the pilots and phase-shifts are designed to enable separate estimation of the channels.

The linear minimum mean-squared error (LMMSE) estimate of $\vect{h}_k$ is given by \cite{Kay1993a}
\vspace{-2mm}
\begin{equation}\label{eq:hk_hat}
\widehat{\vect{h}}_{k} \triangleq \sqrt{K(LR+1)\eta}\overline{\vect{R}}_k^{\mathrm{h}}\left(K(LR+1)\eta\overline{\vect{R}}_k^{\mathrm{h}}+\sigma^2\vect{I}_M\right)^{-1}\vect{z}_{k}^{\mathrm{p,h}}
\vspace{-1.5mm}
\end{equation}
where it follows from \eqref{eq:channel-UE-BS} that $\overline{\vect{R}}_k^{\mathrm{h}}$ can be computed as
\vspace{-2mm}
\begin{equation}
\overline{\vect{R}}_k^{\mathrm{h}}=\mathbb{E}\left\{\vect{h}_k\vect{h}_k^{\Htran}\right\}=\sum_{s=1}^{S_k^{\mathrm{h}}}\bar{\vect{h}}_{k,s}\bar{\vect{h}}_{k,s}^{\Htran}+\vect{R}_k^{\mathrm{h}}. \label{eq:Rhbar}\vspace{-2mm}
\end{equation}

The sufficient statistics for the estimation of $[\vect{H}_{k\ell}]_{:n}$, for $n\in \mathrm{RIS}_{\ell,r}$ is obtained by combining the signals \eqref{eq:sufficient_statistics_k} with appropriate weighting as
\vspace{-2.2mm}
\begin{equation}
\vect{z}_{k,\ell r}^{\mathrm{p,H}}\!=\!\frac{\sum_{t=0}^{LR}\!\psi_{\ell r,t}^{*}\vect{z}_{k,t}^{\mathrm{p}}}{\sqrt{LR+1}}\!=\!\sqrt{\!K\!(LR\!+\!1)\eta}\!\!\!\!\sum_{n\in \mathrm{RIS}_{\ell,r}}\!\!\!\!\left[\vect{H}_{k\ell}\right]_{:n}\!+\!\tilde{\vect{n}}_{k,\ell r}^{\mathrm{p,H}} \label{eq:sufficient_statistics_k_ell}
\vspace{-3.5mm}
\end{equation}
where we have used that ${\bm{\psi}_{\ell^{\prime} r^{\prime}}^{\Htran}\bm{\psi}_{\ell r}=0}$, ${\forall (\ell^{\prime},r^{\prime})\neq (\ell,r)}$ and  ${\bm{\psi}_{\ell r}^{\Htran}{\bf 1}_{LR+1}=0}$,  ${\forall \ell, \forall r}$, and ${\tilde{\vect{n}}_{k,\ell r}^{\mathrm{p,H}}=\sum_{t=0}^{LR}\psi_{\ell r,t}^*\tilde{\vect{n}}_{k,t}^{\mathrm{p}}\big/\sqrt{LR+1} \sim\CN(\vect{0}_M,\sigma^2\vect{I}_{M})}$. 
\vspace{-1.5mm}
\begin{lemma}\label{lemma1}
	The LMMSE estimate of $\left[\vect{H}_{k\ell}\right]_{:n}$, for $n\in \mathrm{RIS}_{\ell,r}$ is given by
	\vspace{-2mm}
	\begin{equation}
	\begin{aligned}[b]
	\left[\widehat{\vect{H}}_{k\ell}\right]_{:n} \triangleq& \sqrt{K(LR+1)\eta}\overline{\vect{R}}_{k\ell,n} \\
	&\times \left(K(LR+1)\eta\overline{\vect{R}}_{k(\ell,r)}
	+\sigma^2\vect{I}_{M}\right)^{-1}\vect{z}_{k,\ell r}^{\mathrm{p,H}} \label{eq:f_kell_estimates}
	\end{aligned}
	\vspace{-3mm}
	\end{equation}
	where 
	\vspace{-3mm}
	\begin{equation}
	\begin{aligned}[b]
	\overline{\vect{R}}_{k\ell,n}=&\sum_{n^{\prime}\in \mathrm{RIS}_{\ell,r}}\left[\overline{\vect{R}}_{k\ell}^{\mathrm{f}}\right]_{nn^{\prime}}\Bigg(\sum_{s=1}^{S_{\ell}^{\mathrm{G}}} \left[\bar{\vect{G}}_{\ell,s}\right]_{:n}\left[\bar{\vect{G}}_{\ell,s}\right]_{:n^{\prime}}^{\Htran}\\
	&+\left[\vect{R}_{\ell}^{\mathrm{G,RIS}}\right]_{n^{\prime}n}\vect{R}_{\ell}^{\mathrm{G,BS}} \Bigg), \\
	\overline{\vect{R}}_{k(\ell,r)} =&\sum_{n\in \mathrm{RIS}_{\ell,r}}\overline{\vect{R}}_{k\ell,n}
	\end{aligned}
		\vspace{-3mm}
	\end{equation}
	with
	$\overline{\vect{R}}_{k\ell}^{\mathrm{f}}=\mathbb{E}\left\{\vect{f}_{k\ell}\vect{f}_{k\ell}^{\Htran}\right\}=\sum_{s=1}^{S_{k\ell}^{\mathrm{f}}}\bar{\vect{f}}_{k\ell,s}\bar{\vect{f}}_{k\ell,s}^{\Htran}+\vect{R}_{k\ell}^{\mathrm{f}}$. 
	\vspace{-2.5mm}
	\begin{proof}
	 The proof is omitted due to space limitation.
	\end{proof}
\end{lemma}
 \vspace{-2.5mm}

As can be seen from Lemma~\ref{lemma1}, we exploit not only the LOS part of the BS-RIS channels $\vect{G}_{\ell}$ but also the other dominant components and the spatial correlation characteristics at both the RIS and BS sides, different from the existing channel estimation schemes (see the introduction for details).

Using the estimated channels, we adjust the RIS phase-shifts for the data transmission according to \eqref{eq:phase-shift1}. Using those phase-shifts, the estimate of the overall channel $\vect{b}_k$ in \eqref{eq:bk} is 
\vspace{-3mm}
\begin{equation} \label{eq:bhat}
\widehat{\vect{b}}_k=\widehat{\vect{h}}_k+\sum_{j=1}^K\widehat{\vect{H}}^{\prime}_{kj}\bm{\phi}_j, \quad k=1,\ldots,K
\vspace{-3mm}
\end{equation}
where the cascaded channel estimates $\widehat{\vect{H}}^{\prime}_{kj}$ can be obtained by picking $[\widehat{\vect{H}}_{k\ell}]_{:n}$ in \eqref{eq:f_kell_estimates} as its columns according to the definition and indexing in \eqref{eq:cascaded}. Once we have obtained $\widehat{\vect{b}}_k$ from \eqref{eq:bhat}, the receive combining schemes given in \eqref{eq:MR} and \eqref{eq:RZF} can be used with the SE expression in Lemma~\ref{eq:capacity1}.

\section{Max-Min Fairness Power Control}
\vspace{-1mm}

In an RIS-assisted massive MIMO system, the most unfortunate UEs with blocked direct channels to the BS or severe pathlosses will be more likely to gain from the phase-shifts introduced by RISs to maximize their signal-to-noise ratios (SNRs). To quantify the improvement over conventional massive MIMO, the worst SEs in the network are thus of great importance. One power control method to emphasize the worst UEs and, hence, UE fairness is the max-min fairness. In this section, we will propose a fixed-point algorithm for the max-min fairness-based power control, where the aim is to maximize the minimum SE among all the UEs (which is equivalent to maximizing the minimum of the SINRs in \eqref{eq:SINR}). The variables of the considered optimization problem are the uplink transmit powers of UEs, i.e., $p_k$, for $k=1,\ldots,K$. The max-min fairness power control problem can be cast as
\vspace{-2.3mm}
\begin{equation}
\begin{aligned}[b]
& \underset{\left\{p_k:k=1,\ldots,K\right\}}{\mathacr{maximize}} \ \ \underset{k\in\{1,\ldots,K\}}{\mathacr{min}}  \\
&  \frac{ p_{k} \left |\mathbb{E} \left\{  \vect{v}_{k}^{\Htran} \vect{b}_k \right\}\right|^2  }{ 
	\sum\limits_{i=1}^K p_{i}  \mathbb{E} \left\{| \vect{v}_{k}^{\Htran} \vect{b}_{i} |^2\right\}
	- p_{k} \left |\mathbb{E} \left\{  \vect{v}_{k}^{\Htran} \vect{b}_{k} \right\}\right|^2  + \sigma^2 \mathbb{E} \left\{\|  \vect{v}_{k} \|^2\right\}
}  \\
& {\mathacr{subject \ to}} \quad  0\leq p_k\leq p_{\rm max}, \ \   k=1, \ldots, K \label{eq:constraint-maxmin}
\end{aligned}
\vspace{-2.3mm}
\end{equation}
where $p_{\rm max}$ is the maximum uplink data transmission power for each UE. We will use the following lemma to solve the above problem optimally using a simple fixed-point algorithm.
\vspace{-5mm}
\begin{lemma} \label{lemma:convergence}
	The fixed-point algorithm whose steps are outlined in Algorithm~\ref{alg:fixed-point} converges to the optimal solution of \eqref{eq:constraint-maxmin}.
\end{lemma}
\vspace{-4mm}
\begin{proof}
	The proof follows from \cite[Lem.~1, The.~1]{Hong2014}.
\end{proof} 
\vspace{-4mm}

\begin{algorithm}[H]
	\caption{Fixed-point algorithm for solving the max-min fairness problem in \eqref{eq:constraint-maxmin}.} \label{alg:fixed-point}
	\begin{algorithmic}[1]
		\State {\bf Initialization:} Set arbitrary $p_k>0$, for $k=1,\ldots,K$, and the solution accuracy $\varepsilon>0$.
		\While{$\underset{k\in\{1,\ldots,K\}}{\max} \textrm{SINR}_k-\underset{k\in\{1,\ldots,K\}}{\min} \textrm{SINR}_k > \varepsilon$} 
		\State $p_k \gets \frac{ \sum\limits_{i=1}^K p_{i}  \mathbb{E} \left\{| \vect{v}_{k}^{\Htran} \vect{b}_{i} |^2\right\}
			- p_{k} \left |\mathbb{E} \left\{  \vect{v}_{k}^{\Htran} \vect{b}_{k} \right\}\right|^2  + \sigma^2 \mathbb{E} \left\{\|  \vect{v}_{k} \|^2\right\}}{\left |\mathbb{E} \left\{  \vect{v}_{k}^{\Htran} \vect{b}_k \right\}\right|^2  }$, $\quad$ $k=1,\ldots,K.$
		\State $\widetilde{p} \gets \underset{k\in\{1,\ldots,K\}}{\max}p_k$.
		\State $p_k \gets \frac{p_{\rm max}}{\widetilde{p}}p_k$, $\quad$ $k=1,\ldots,K.$
		\EndWhile
		\State {\bf Output:} $p_1,\ldots,p_K$.
	\end{algorithmic}
	\end{algorithm}
\setlength{\textfloatsep}{0.1cm}
\setlength{\floatsep}{0.1cm}

\vspace{-3mm}
Note that the BS is responsible for the computations regarding the channel estimation, phase-shifts selection, and the power control, which have all relatively low complexity due to the closed-form equations with basic linear algebra.

\vspace{-1mm}

\section{Numerical Results and Discussion}
\vspace{-1.3mm}

In this section, we quantify the performance gain of the proposed RIS-assisted massive MIMO compared to conventional massive MIMO. The pathloss models and the shadow fading parameters for the LOS and NLOS paths originate from \cite[Table~5.1]{channel} for an urban microcell environment. 
Isotropic antennas are considered at the BS and UEs. The RIS elements have area $(\lambda/4)^2$, where $\lambda$ is the wavelength, and are deployed with $\lambda/4$ spacing. The carrier frequency is $1.9$\,GHz and the noise figure is $7$\,dB.  The maximum uplink power for each UE is $100$\,mW per $1$\,MHz bandwidth and each UE transmits with this power during pilot transmission. In the data transmission phase, the proposed max-min fair power control is adopted. The number of BS antennas is $M=100$ and they are deployed as a half-wavelength-spaced uniform linear array (ULA). There are $L=2$ RISs, each being a uniform planar array (UPA) with $16\times16$ elements. The spatial correlation matrices for both the BS and RISs are generated using the 3D Gaussian local scattering model for UPAs and ULAs with 15 degrees angular spread \cite[Sec.~7.3.2]{massivemimobook}.\footnote{For complexity reasons, the correlation matrices were tightly approximated similar to the small angular deviation assumption in \cite[Sec.~2.6]{massivemimobook}.} Each RIS is assigned to the one of the two UEs with the lowest BS-UE channel gains. The coherence block length is $\tau_c=10\,000$ samples and  $\tau_p=20K$ mutually orthogonal pilot sequences are used for conventional massive MIMO.\footnote{It was observed that for the considered scenario, this selection for $\tau_p$ presents a good trade-off between pilot signal SNR and data transmission length in each coherence block.} For the RIS-assisted massive MIMO case, each $4\times4$ set of RIS elements is reconfigured to have the same phase-shift during pilot transmission and, hence, we have $\tau_p=(LR+1)K=33K$.

We assume the RISs are deployed to always have LOS paths to the BS. The existence of the LOS for the other channels is modeled in a probabilistic manner and the formulas for it and the Ricean $\mathcal{K}$-factor are given in \cite[Sec.~5.5-3]{channel}, unless otherwise stated. The BS and the RISs are assumed to be mounted $10$\,m above the height of the UEs. The 2D locations of the BS and two RISs is $(0,0)$, $(10,30)$, and $(10,-30)$, respectively, where the unit is meters. In each setup, $K=10$ cell-edge UEs are randomly dropped in the $100$\,m $\times\ 50\,$m area that extends from $(200,-25)$ to $(300,25)$.

\begin{figure}[t]
	\includegraphics[trim={1.6cm 0.1cm 2.6cm 0.8cm},clip,width=3.3in]{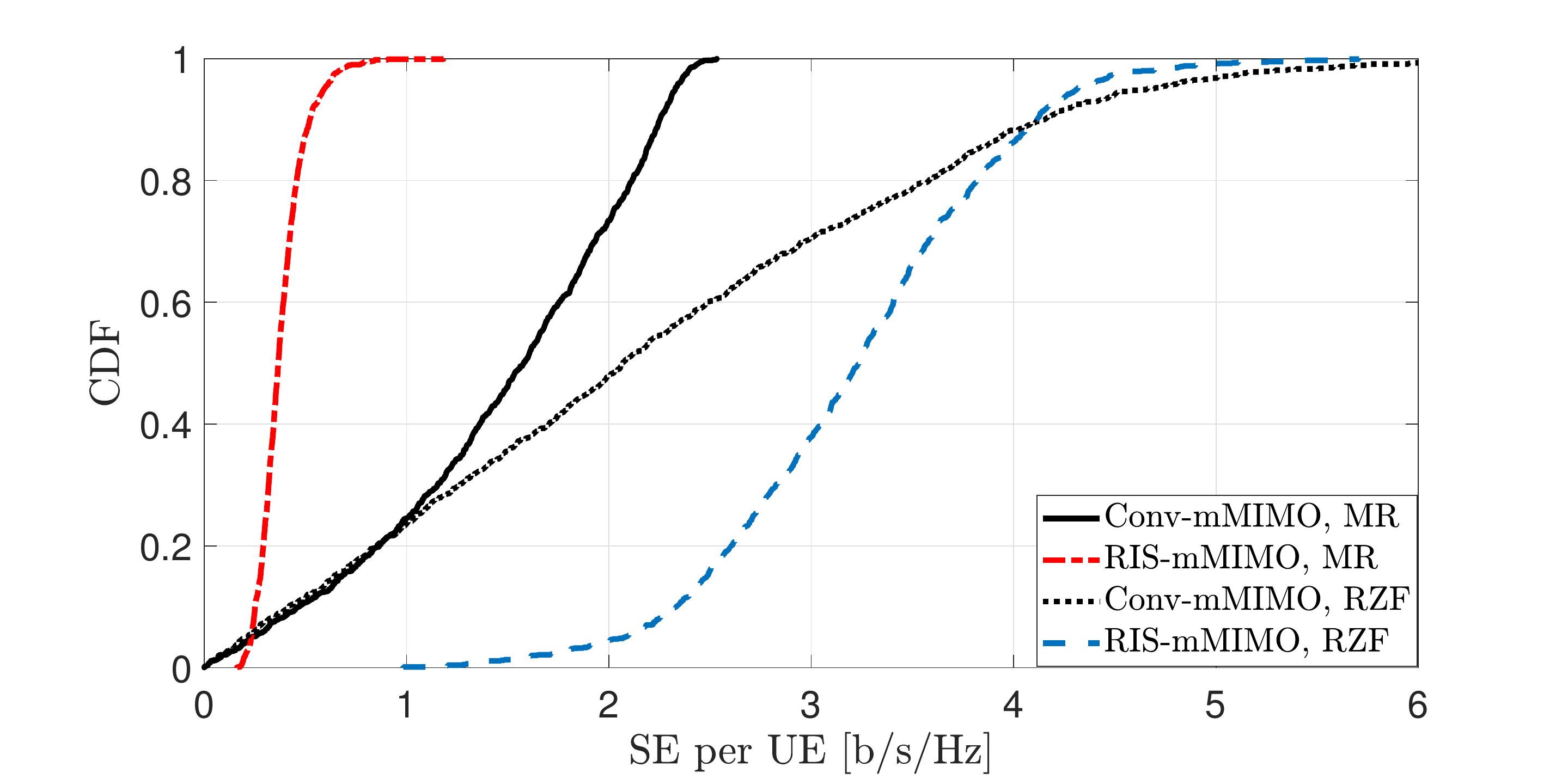}
	\vspace{-0.4cm}
	\caption{CDF of SE per UE for conventional and RIS-assisted massive MIMO with different receive combiners.}
	\label{fig:fig1}
	\vspace{-0.1cm}
\end{figure}

In Fig.~\ref{fig:fig1}, we plot the cumulative distribution function (CDF) of the SE per UE for both conventional massive MIMO (Conv-mMIMO) and the proposed RIS-assisted massive MIMO (RIS-mMIMO) when using either MR or RZF receive combining. The randomness is the UE locations. There is always an LOS path for the channel between any RIS and UE. For all the channels, the LOS path is the only specular component once LOS exists. For MR combining, RIS-mMIMO starts from a higher SE in the lower tail of the CDF curve than with Conv-mMIMO, which indicates that the RISs effectively improves the SE of very unfortunate UEs. However, this comes with a significant SE reduction for other UEs. The reason is that the RISs are configured to improve for the most unfortunate UEs, which creates additional scatterers compared to non-RIS case. This creates additional interference which cannot be suppressed using MR combining in an RIS-assisted massive MIMO system.
On the other hand,  RIS-mMIMO provides a significant SE improvement when RZF combining is used to reject interference. The median SE (the point where the CDF is 0.5) provided by RIS-mMIMO is 1.5 times larger compared to Conv-mMIMO. Moreover, the 90\%-likely SE (the point where the CDF is 0.1) is approximately 5.6 times greater with RIS-mMIMO. The lower tails of the CDF curves are of great importance since low-rate UEs  need to be active more often than high-rate UEs, when using the same service.

\begin{figure}[t]
	\includegraphics[trim={1.6cm 0.1cm 2.6cm 0.8cm},clip,width=3.3in]{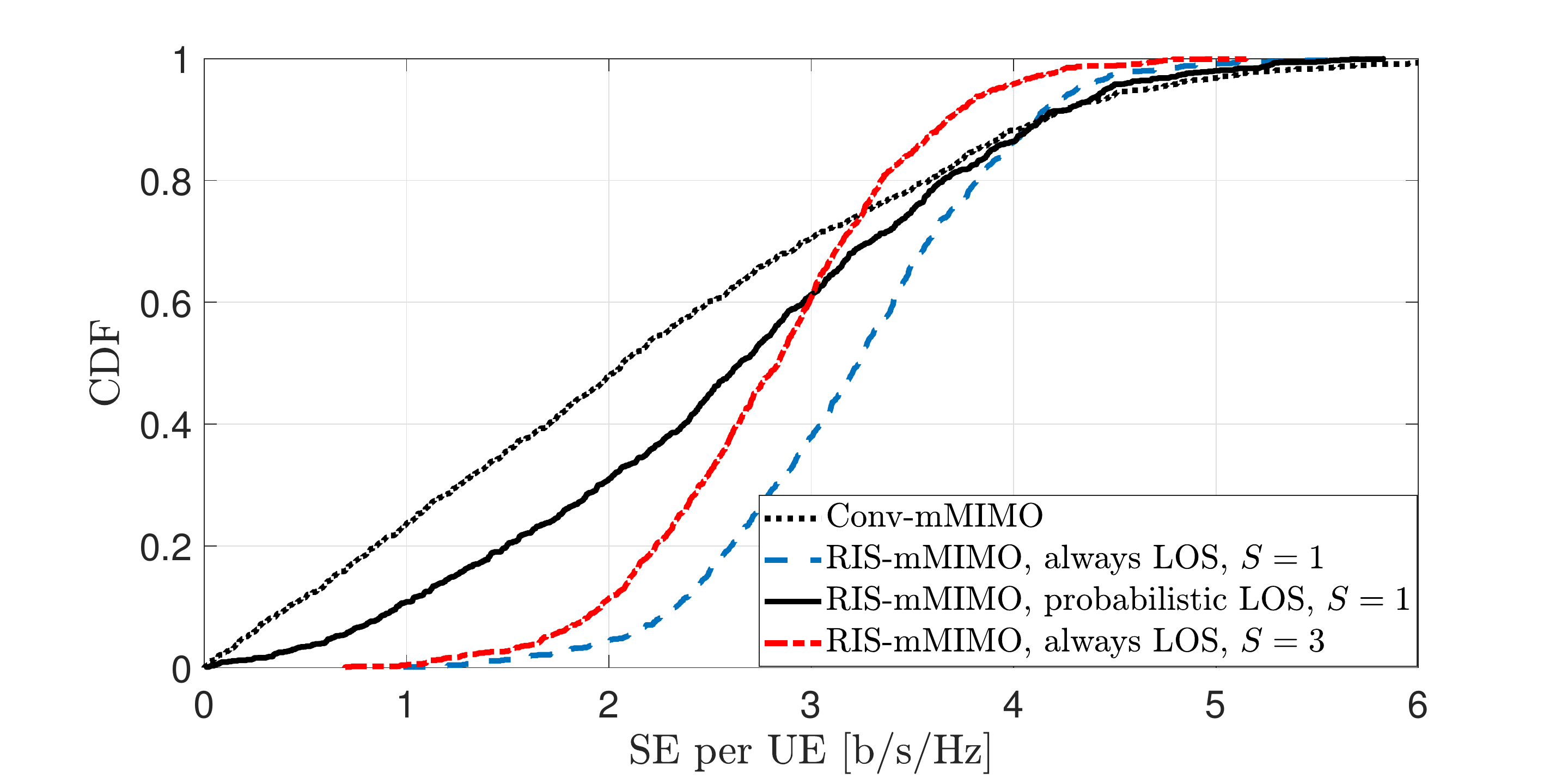}
	\vspace{-0.4cm}
	\caption{CDF of SE per UE for conventional and RIS-assisted massive MIMO with different fading scenarios and RZF combiner.}
	\label{fig:fig2}
	\vspace{-0cm}
\end{figure}

To see the impact of the LOS and other specular components, we reconsider the RZF curves from Fig.~\ref{fig:fig1} in Fig.~\ref{fig:fig2} and compare them with RIS-mMIMO under  different propagation conditions. The CDF of RIS-mMIMO from Fig.~\ref{fig:fig1} is denoted by ``always LOS, $S=1$'' in Fig.~\ref{fig:fig2} since there is always an LOS path between any RIS and UE, and the number of specular components of any channel is $S=1$ once LOS exists. The ``probabilistic LOS'' case corresponds to the scenario when the LOS probabilities for the RIS-UE channels are determined according to \cite[Sec.~5.5-3]{channel}. We also include the scenario where there is always an LOS path for any RIS-UE channel and the number of specular components is $S=3$ for the BS-RIS and RIS-UE channels. In this case, the original LOS channel gain is distributed randomly over two non-LOS dominant components by keeping the power ratio of the LOS component being 0.5. As demonstrated in Fig.~\ref{fig:fig2}, when the LOS is probabilistic, the performance improvement RIS-mMIMO provides over Conv-mMIMO becomes less apparent compared to the ``always LOS'' case. Apart from the LOS path existence, there is an important factor that leads to this result. According to the model in \cite{channel}, the channel gain is higher when LOS exists compared to the NLOS case. This effect is observed from the last plot for the $S=3$ and ``always LOS'' case. Although the power is distributed among three specular components, RIS-mMIMO provides much higher SE for the unfortunate UEs as can be seen from the lower tails. However, $S=1$ and ``always LOS'' case is the most preferable scenario. 

An alternative to the proposed LMMSE channel estimation method is the least squares (LS) estimator that can be applied from \cite{RISchannelEstimation_nested_cascaded, RISchannelEstimation_nested_LS_single}  to our problem setup. However, due the the pilot contamination resulting from the shared phase-shifts, we have observed the SEs obtained by LS estimator are significantly lower in comparison to LMMSE estimator.

\vspace{-0.5mm}
\section{Conclusion}
\vspace{-0.5mm}

We have proposed an LMMSE channel estimation scheme with low training overhead and the respective phase-shift and receive combiner design for an RIS-aided massive MIMO system in a multi-specular spatially correlated fading environment. The simulation results demonstrate that the cell-edge UEs, in particular the most unfortunate UEs, can benefit a lot from the assistance of the RISs when RZF combiner is used. On the other hand, for MR combiner using RIS does not improve conventional massive MIMO. When there is always an LOS between any RIS and UE and there is no other dominant component, the SE improvement is higher in comparison to probabilistic LOS and multi-specular fading.

\bibliographystyle{IEEEtran}
\bibliography{IEEEabrv,refs}

\end{document}